\documentclass[aps]{revtex4}
\usepackage{amsmath}
\usepackage{graphicx}
\begin{document}
\title{Stellar matter in the Quark-Meson-Coupling Model with neutrino trapping}
\author{P.K. Panda}
\author{D.P. Menezes}
\affiliation{Depto de F\'isica-CFM, Universidade Federal de Santa
Catarina, CP. 476, 88040-900 Florian\'opolis-SC, Brazil}
\author{C. Provid\^encia}
\affiliation{Centro de F\'isica Te\'orica - Dep. de F\'isica,
Universidade de Coimbra, P-3004 - 516 Coimbra, Portugal}
\begin{abstract}
The properties of hybrid stars formed by hadronic and quark matter
in $\beta$-equilibrium are described by appropriate 
equations of state (EoS) in the framework of the quark meson coupling (QMC) 
model. In the present  work we include the possibility of trapped neutrinos 
in the equation of state and obtain the properties of the related hybrid stars. 
We use the quark meson coupling model for the hadron matter and two 
possibilities for the quark matter phase, namely, the unpaired quark phase
and the color-flavor locked phase. The differences are discussed and a 
comparison with other relativistic EoS is done. 

PACS number(s): {95.30.Tg, 21.65.+f, 12.39.Ba, 21.30.-x}
\end{abstract}
\maketitle
During the early stage of a proto-neutron star neutrinos get trapped in it
when their mean-free path is smaller than the star radius. The presence of 
neutrinos generally gives rise to a stiffer EoS. This may have 
important consequences on the evolution of the star,
namely, it can happen that during the cooling process the star decays into a 
low-mass black hole \cite{prak97}.
It is worth mentioning that over 20 years ago the importance of neutrino 
trapping was already pointed out in \cite{gb}, where the authors claimed that
another important effect of including trapped neutrinos is that the collapse 
is gentler in its presence than it would be without it.

In the present paper we are interested in building the neutrino trapped
EoS for mixed matter of quark and hadron phases. We 
employ the QMC model (QMC)~\cite{guichon,ST,pal}
in order to describe the hadron phase. For the quark phase we use
two distinct models, the unpaired quark model (UQM), which is given by the 
simple MIT bag model \cite{bag} and the color-flavor locked phase (CFL) 
\cite{raja,ar} in which quarks are paired near the Fermi surface  forming 
a superconducting phase \cite{bl}.

In a previous work \cite{recente} we have used the same formalism in order to 
study the properties of hybrid stars at $T=0$ MeV. 
In this work, we verify the effect of including trapped neutrinos in the 
same spirit as done in \cite{mp}, where we have seen that the EoS changes 
considerably and the mixed phase appears at higher energy densities than in 
the EoS built without the inclusion of neutrinos in accordance with what 
was seen in other works, as in \cite{prak97} for instance.
We have also verified that 
trapping keeps the electron population high so that dense matter contains more
protons (and depending on the parametrization used, other positively charged
particles) than matter without neutrinos. This fact was also discussed in 
\cite{vbpr}. 

Proto-neutron stars with a certain baryonic mass at birth keep this mass during
its evolution until the final neutrino free star because most of its matter is
accreted in the early stages after birth. Hence, stars with trapped neutrinos 
and a baryonic mass larger than the corresponding ones after deleptonization
collapse to a black hole. In \cite{mp} we have described the hybrid star 
within a non-linear Walecka model (NLWM) \cite{Glen00} including the baryonic 
octet and a phase transition to a quark phase. Within this description  
the maximum baryonic masses supported by neutrino trapped EoSs are 
larger than the corresponding ones found in neutrino free EoSs, a fact which 
occurs in other EoS which include the strangeness degree of freedom 
\cite{prak97,vbpr}. In what follows we investigate whether this behavior 
is also present in the framework of the QMC model.
Because of its importance in understanding the evolution process in a star, 
we  calculate the baryonic masses for the hybrid stars studied in the 
present work. We also compare the properties of the stars obtained within the 
UQM model and the CFL model.

We perform all the calculations at $T=0$ MeV although the neutrino 
trapped phase occurs for temperatures  in the interior of the star which can 
vary between 20-40 MeV \cite{prak97,mp}. However, it was shown in \cite{mp} 
that the effect of trapping is much stronger than the
finite temperature effect. Therefore, we believe that the main conclusions 
drawn in the present work are still valid for a finite temperature 
calculation.  On the other hand, the calculation with the
CFL phase will give us only an upper limit, since at finite temperature there 
is a phase transition  from the color superconducting state to a normal phase, 
which,  according to \cite{rischke},  is a second order transition with a BCS 
critical temperature $T_c\sim 0.57 \Delta$, $\Delta$ being the gap parameter.

Since most of the analytical calculations and formulae used in the present 
work have already been given in \cite{recente} they  will be omitted here.
As mentioned above, we have used the QMC model with the inclusion of hyperons
for the hadron phase. In this model, the nucleon in nuclear medium is assumed 
to be a static spherical MIT bag in which quarks interact with the scalar and
vector fields, $\sigma$, $\omega$ and $\rho$ and these
fields are treated as classical fields in the mean field
approximation.  The quark field, $\psi_q(x)$, inside the bag then
satisfies the equation of motion:
\begin{eqnarray}
&&\left[i\,\rlap{/}\partial-(m_q^0-g_\sigma^q\, \sigma)
-g_\omega^q\, \omega\,\gamma^0 + \frac{1}{2} g^q_\rho \tau_z \rho_{03}\right]
\psi_q(x)=0\nonumber\\
&&\hspace {2.2in} q=u,d,s,
\label{eq-motion}
\end{eqnarray}
where $m_q^0$ is the current quark mass, and $g_\sigma^q$,
$g_\omega^q$ and $g_\rho^q$ denote the quark-meson coupling constants.
After enforcing the boundary condition at the bag surface, 
the transcendental equation for the ground state solution of the quark (in
$s$-state) is
$
j_0(x_q)=\beta_q j_1(x_q)
$
which determines the bag eigenfrequency $x_q$, where
$\beta_q=\sqrt{(\Omega_q-R_B m_q^*)/(\Omega_q+R_B m_q^*)}$, 
with $\Omega_q=(x^2_q+R^2{m^*_q}^2)^{1/2}$;
$m^*_q=m^0_q-g_\sigma^q\sigma_0$, is the effective quark mass.
The energy of the nucleon bag is
$M^*_B=3\frac{\Omega_q}{R_B}-\frac{Z_B}{R_B}+\frac{4}{3}\pi R^3_B B_B$,
where $B_B$ is the bag constant and $Z_B$ parameterizes the sum of the 
center-of-mass motion and the gluonic corrections. 
The bag radius, $R_B$, is then obtained through
the stability condition for the bag.
An interesting fact related to the QMC model is that the bag volume changes
in the medium through the mean value of the $\sigma$-field. This also implies
that the bag eigenvalues are also modified.
The onset of hyperons depends on the conditions of chemical equilibrium 
and charge neutrality discussed below and also on the meson-hyperon coupling
constants for which we have chosen
the hyperon coupling constants constrained by the binding of the 
$\Lambda$ hyperon in nuclear matter, hyper-nuclear levels and neutron star 
masses ($x_\sigma=0.7$ and $x_\omega=x_\rho=0.783$) and assumed that the 
couplings to the $\Sigma$ and $\Xi$ are equal to those of the $\Lambda$ 
hyperon \cite{Glen00}. 
The leptons either in the hadron or in the quark phase are considered as free
Fermi gases and they do not couple with the hadrons, with the mesons or with 
the quarks. 

The condition of chemical equilibrium is  imposed through the  two 
independent chemical potentials for neutrons $\mu_n$ and electrons  
$\mu_e$ and it implies that the chemical potential of baryon $B_i$ is 
$\mu_{B_i}=Q_i^B\mu_n-Q^e_i \mu_e$, where $Q^e_i$ and $Q_i^B$ are,
respectively, the electric and baryonic charge of baryon or quark $i$.
Charge neutrality implies $\sum_{B_i} Q^e_i \rho_{B_i} + \sum_l q_l \rho_l=0$ 
where $q_l$ stands for the electric charges of leptons. 
If neutrino trapping is imposed to the system, the beta equilibrium condition
is altered to $\mu_{B_i}=Q_i^B\mu_n-Q^e_i (\mu_e-\mu_{\nu_e})$.
In this work we have not included trapped muon neutrinos. Because of
the imposition of trapping the total leptonic number is conserved, i.e.,
$Y_L=Y_e+Y_{\nu_e}=0.4$. 

For the quark phase we consider two models. First of all, we take the quark 
matter EoS as in \cite{bag} in which $u$,$d$ and $s$
quark degrees of freedom are included in addition to electrons.
Up and down quark masses are set to zero and the strange quark mass is 
taken to be either 150 or 200 MeV so that we are able check the effect of the 
$s$-quark mass. In chemical
equilibrium $\mu_d=\mu_s=\mu_u+\mu_e$. In terms of neutron and
electric charge chemical potentials $\mu_n$ and $\mu_e$, one has
$
\mu_u={1\over 3}\mu_n-{2\over 3}\mu_e,\quad
\mu_d={1\over 3}\mu_n+{1\over 3}\mu_e,\quad
\mu_s={1\over 3}\mu_n+{1\over 3}\mu_e.$
In the energy density for the quark matter EoS a term $+B$ and in the 
pressure a 
term $-B$ are inserted. This term is responsible for the simulation of 
confinement. For the Bag model, we have taken B$^{1/4}$=190 and 200 MeV.

In the EoS taking into consideration a CFL quark paired phase, 
the quark matter is treated as a Fermi sea of free quarks
with an additional contribution to the pressure arising from the 
formation of the CFL condensates. The density of the three types of quarks
are identical and the electron density is zero, as shown in \cite{raja}.
The expressions for the energy density and
pressure depend on a gap parameter $\Delta$ which is taken to be 100 MeV 
\cite{ar}. 

Once the hadron and quark phases are well established, we have to construct 
the mixed phase, imposing charge neutrality  globally, 
$\chi\, \rho_c^{QP}+ (1-\chi) \rho_c^{HP}+\rho_c^l=0$,
where $\rho_c^{iP}$ is the charge density of the phase $i$, $\chi$ is the 
volume fraction occupied by the quark phase and $\rho_c^l$ is the electric 
charge density of leptons. We consider a uniform background of leptons in the 
mixed phase since Coulomb interaction has not been taken into account. 
According to the Gibbs conditions for phase coexistence, the baryon chemical 
potentials, temperatures and pressures have to be identical in both phases, 
i.e., $\mu_{HP,n}=\mu_{QP,n}=\mu_n, \quad \mu_{HP,e}=\mu_{QP,e}=\mu_e, 
\quad T_{HP}=T_{QP},\quad P_{HP}(\mu_n,\mu_eT)=P_{QP}(\mu_n,\mu_e,T),$
reflecting the needs of chemical, thermal and mechanical equilibrium, 
respectively.

\begin{figure}
\begin{tabular}{cc}
\includegraphics[width=5.cm]{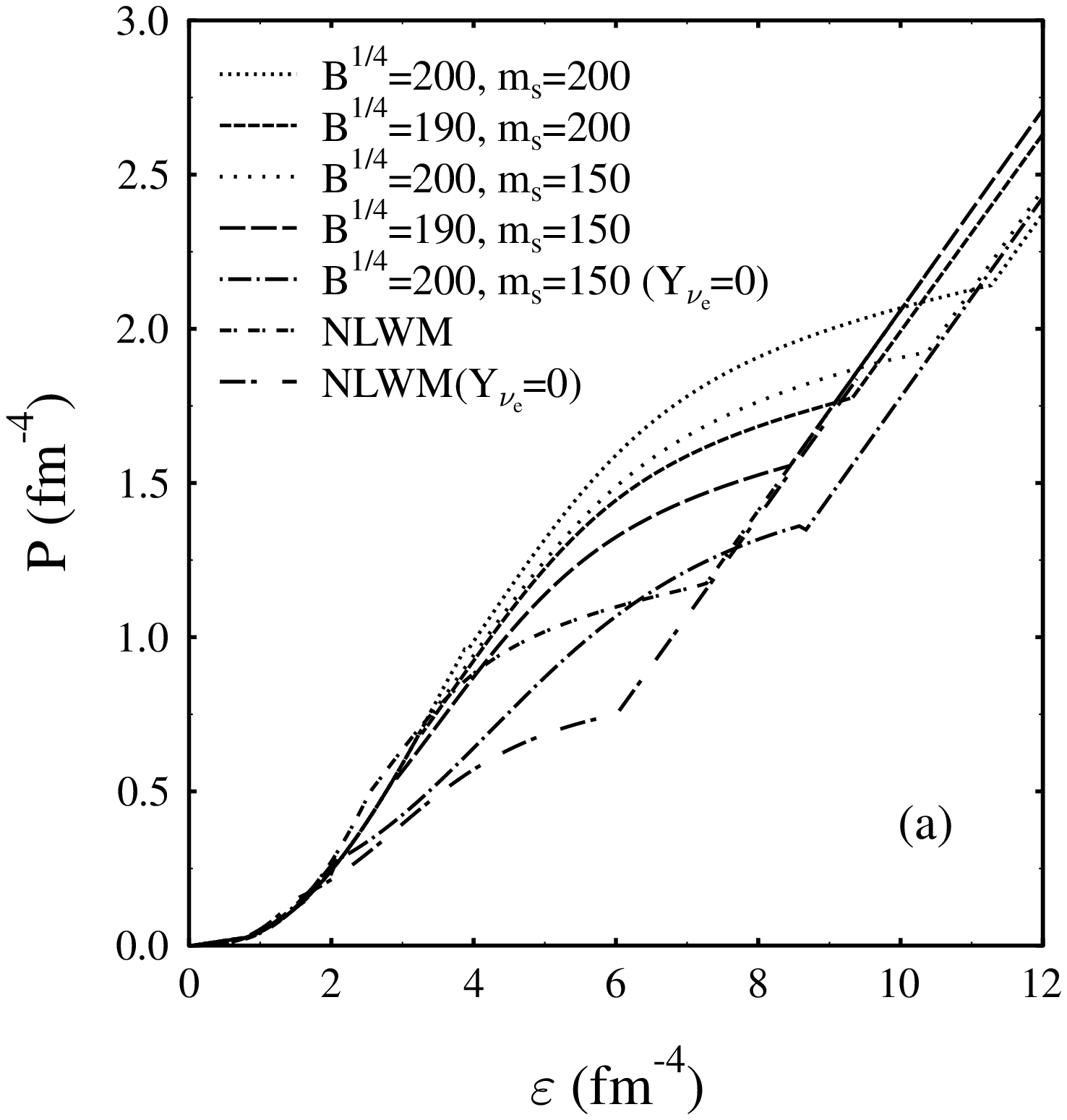} &
\includegraphics[width=5.cm]{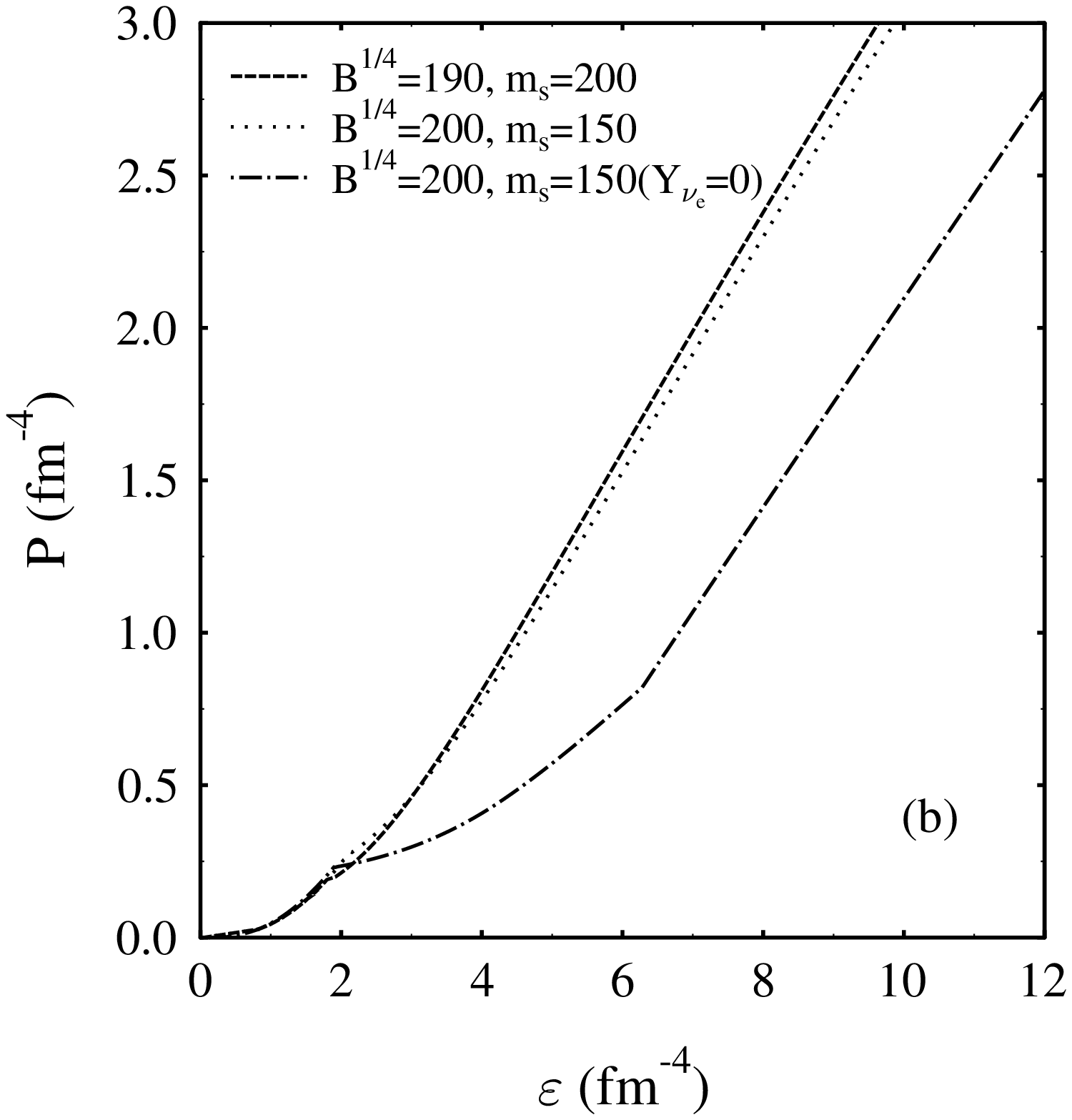}
\end{tabular}
\caption{Equation of state obtained with the QMC model plus (a) UQM (b) CFL.}
\label{eos}
\end{figure}

In fig. \ref{eos}, the EoSs obtained with both quark models are displayed for
different $B$ values and two strange quark masses with neutrino trapping  
($Y_L=0.4$). For the sake of comparison we have also plotted one EoS with
no neutrinos ($Y_{\nu_e}=0$), and included the EoS obtained within a NLWM 
formalism for the hadronic phase \cite{mp} and an UQM with $m_s=150$ MeV and 
$B^{1/4}=190$MeV, with and without neutrino trapping. As already discussed 
in \cite{vbpr,prak97} 
the EoSs are harder if neutrino trapping is imposed, independently of the 
model used.  A larger $s$-quark mass and a larger $B$ parameter make the 
quark  EoSs harder in the mixed phase, a fact that
manifests itself on the maximum mass stellar configuration. The main 
differences between the QMC formalism and the NLWM are: a) the NLWM  EoS is 
harder at low densities and softer at intermediate densities  due to the
presence of hyperons; b) the transition to a pure  quark phase occurs at 
lower densities in the NWLM. This behavior has consequences on the 
properties of the corresponding  families of stars. 

\begin{figure}
\begin{tabular}{cc}
\includegraphics[width=5.0cm,angle=0]{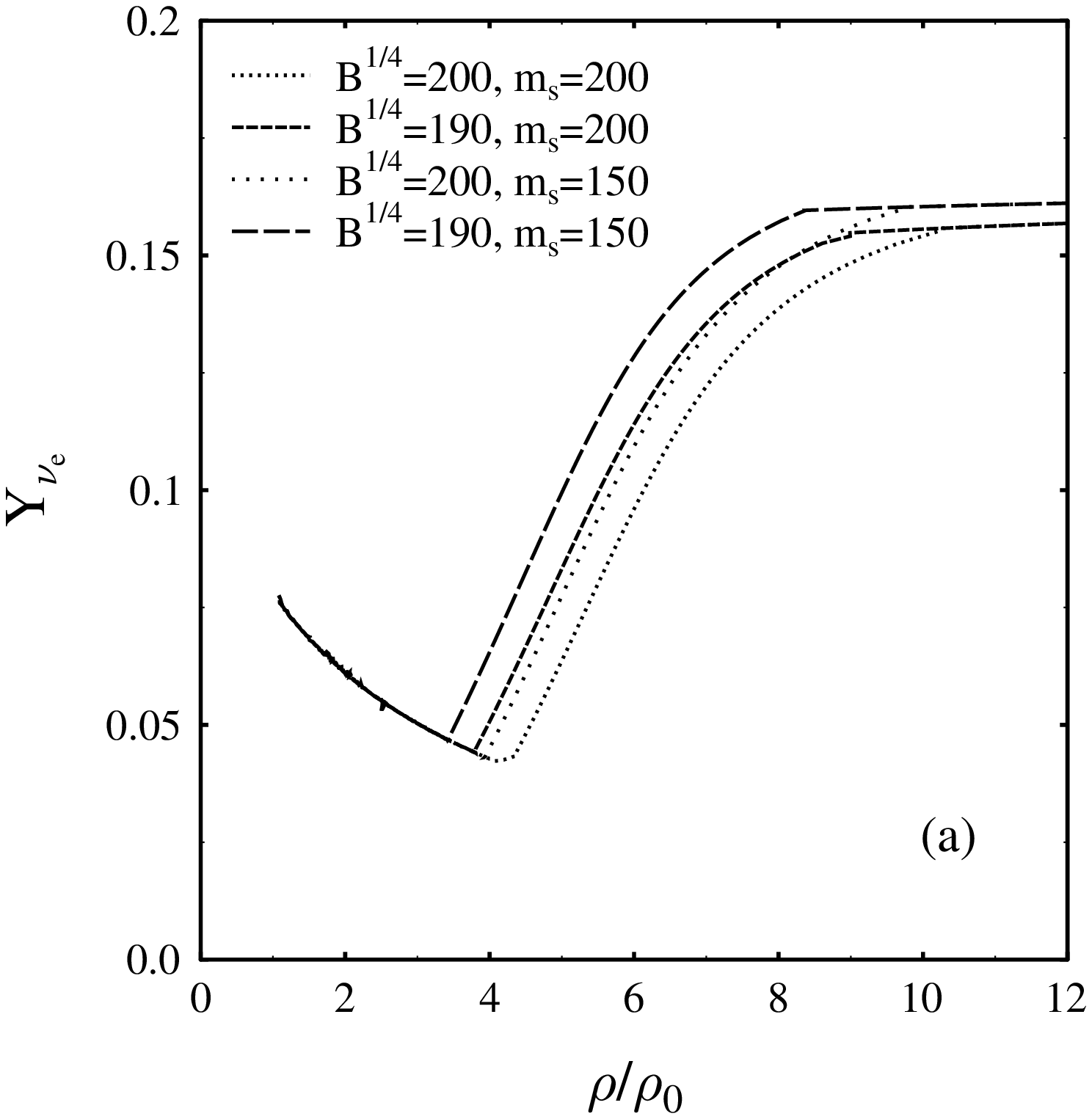}&
\includegraphics[width=5.0cm,angle=0]{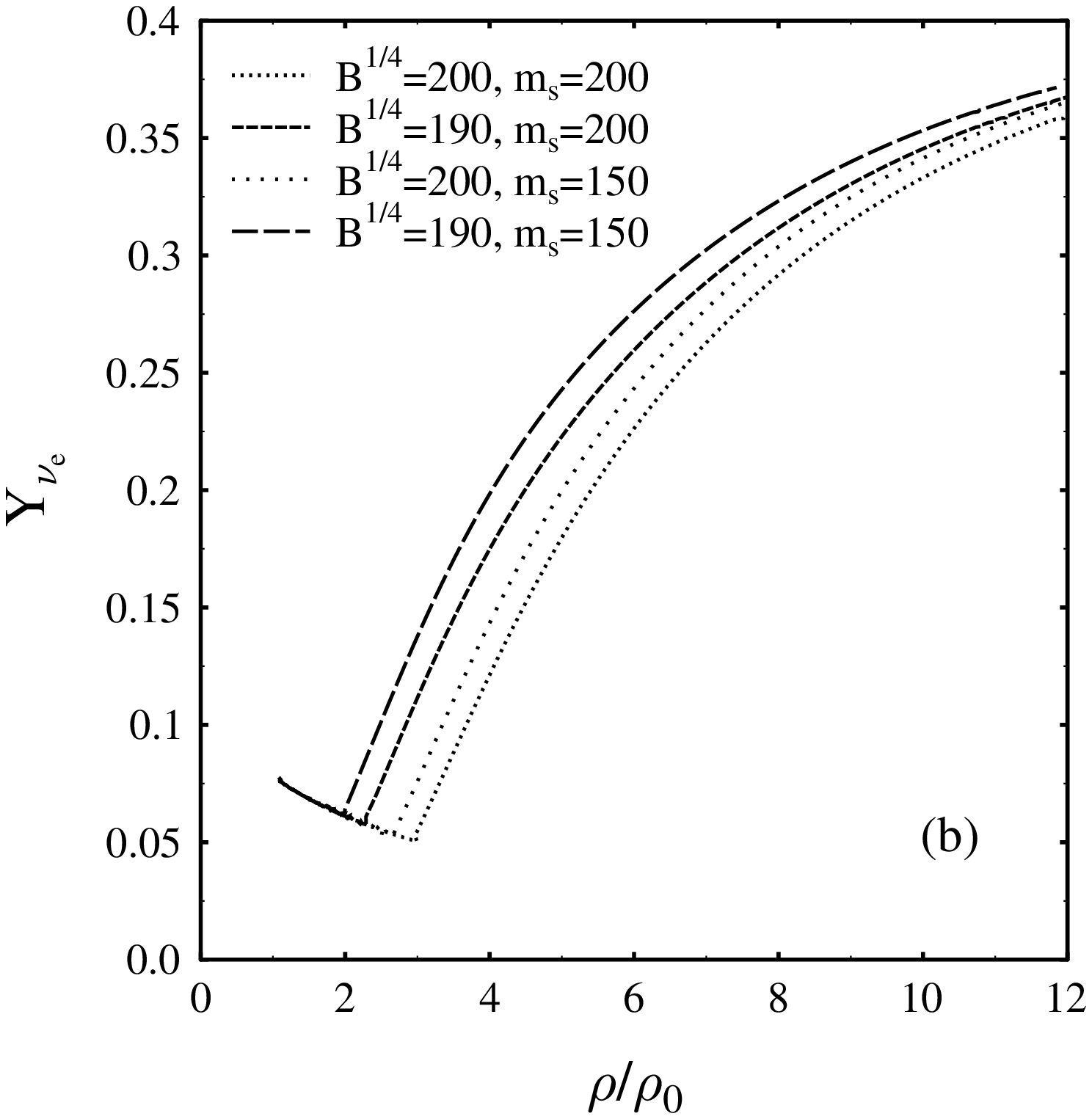}
\end{tabular}
\caption{Neutrino fraction for the EoS obtained with the QMC model plus (a) UQM 
(b) CFL.}
\label{neutrinof}
\end{figure}

In order to better understand the importance of the neutrinos when neutrino 
trapping is imposed, in fig. \ref{neutrinof} the fraction of
neutrinos is shown. The behavior encountered for the neutrino fraction 
if the UQM is used resembles the one shown in \cite{mp}: the population 
of neutrinos decreases in the hadron phase and, contrary to \cite{mp}, 
only  increases in the mixed and 
quark phases. The highest yields are of the order of 0.16. In this model, 
for smaller values of the 
strange quark mass, the neutrino population at high densities is greater.
This occurs because a smaller $s$-quark mass gives larger $s$-quark densities 
and therefore, a smaller electron fraction. A fixed lepton fraction then 
implies a larger $\nu_e$ fraction.

If the CFL is chosen for the quark phase, the population of neutrino is
higher in the mixed and quark phases. This is due to the fact that in the CFL 
phase no electrons are present since the number of $u$, $d$ and $s$ quarks 
are equal, therefore the lepton fraction is kept constant only by the presence 
of neutrinos. This implies a much greater neutrino flux during 
deleptonization. 
However, in a self-consistent calculation at finite temperature we do not 
expect such a strong effect  since pairing will be weaker.

The amount of neutrinos depends on the fraction of charged  hyperons  and 
quarks present in each phase, which are determined by the models used and 
consequently by the bag pressure and the strange quark mass. In the present
approach with the CFL phase or  UQM for $B^{1/4}=190$ MeV  the onset
of   hyperons, if any,  occurs for $\rho>10\rho_0$. With the UQM and 
$B^{1/4}=200$ MeV the   hyperon onset occurs at $\sim 4 \rho_0$ but the 
charged hyperon fraction, namely $\Sigma^-$,  is never larger than 0.004.
\begin{table}
\begin{ruledtabular}
\caption{Star properties}
\label{tab2}
\begin{tabular}{lcccccc}
model & $B^{1/4}$&$m_s$& $\frac{M_{max}}{M_\odot}$ & 
$\frac{M_b}{M_\odot}$ & R& $\varepsilon_0$ \\
& (MeV)&(MeV) &  & 
& (km)& (fm$^{-4}$) \\
\hline
QMC+UQM & 190& 150 &1.94 & 2.15 & 12.09 & 5.47 \\
QMC+UQM & 190& 200 &1.98 & 2.20 & 12.02 & 5.59 \\
QMC+UQM & 200& 150 &1.99 & 2.22 & 11.97 & 5.63 \\
QMC+UQM & 200& 200 &2.02 & 2.26 & 11.89 & 5.77 \\
\hline
QMC+UQM & 200& 150 &1.73 & 1.93 & 12.44 & 4.89 \\
($Y_{\nu_e}=0$)& &  & &  &  &  \\
\hline
NLWM+UQM & 190 & 150 & 1.64& 1.83 & 12.84 & 4.5\\
($Y_{\nu_e}=0$)\cite{mp}& &  & &  &  & \\
NLWM+UQM & 190&150&2.00& 2.22& 12.59 & 5.06\\
($Y_L=0.4$)\cite{mp}& &&& &  & \\
\hline
QMC+CFL & 190& 150 &1.80 & 1.99 & 10.86 & 7.32 \\
QMC+CFL & 190& 200 &1.84 & 2.03 & 11.10 & 6.97 \\
QMC+CFL & 200& 150 &1.83 & 2.02 & 11.36 & 6.57 \\
QMC+CFL & 200& 200 &1.87 & 2.07 & 11.57 & 6.20 \\
\hline
QMC+CFL & 200& 150 &1.49 & 1.65 & 13.4 & 3.32 \\
($Y_{\nu_e}=0$) & & & &  & &  \\
\end{tabular}
\end{ruledtabular}
\end{table}
Hybrid neutron star profiles can be obtained from all the EoS 
studied by solving the Tolman-Oppenheimer-Volkoff equations. 
Even at finite temperature the conditions of hydrostatic equilibrium are 
nearly fulfilled \cite{bl86}. In Table I we show the values obtained for the 
maximum gravitational and  baryonic masses and
radii of neutron stars as function of the central density for the EoSs 
studied in this work. For a fixed  bag constant, the stellar and baryonic 
masses of the most massive stable stars are higher for higher strange quark 
masses in both models.
For the UQM, these critical  masses are always higher than for the CFL model, 
because it also corresponds to harder EoS.
The radii and central energy density depend on the
model and on the strange quark mass.
The radii values are larger for the UQM model and the central
energy density are larger for the CFL model, again due to the fact that the 
UQM EoSs are harder. Comparing the results of
table I with the ones presented in \cite{recente}, where neutrino trapping was 
not considered, one can see that the inclusion of trapping makes the
gravitational masses reasonably higher. The same conclusion was drawn in 
\cite{mp} where the calculations were performed with different relativistic 
models. In the Table I we have also included properties of maximum mass  stars 
obtained within the NLWM for the hadronic phase \cite{mp}. Two conclusions are 
in order:  the  maximum baryonic masses obtained within the NLWM are larger 
and the difference of maximum masses for trapped and
untrapped matter is smaller for the QMC ($\sim 0.2 M_\odot$) than for the 
NLWM  ($\sim 0.4 M_\odot$).  This means that the number of stars that would 
decay into a blackhole is much smaller in the QMC model and is probably due 
to the fact that no hyperons are formed in the interior of stars obtained 
with QMC for $m_s=150$ MeV and $B^{1/4}=190$ MeV contrary to the
NLWM case. If the quark phase is a CFL state the baryonic mass difference 
between the neutrino rich stars and neutrino poor is greater than in the UQM, 
($\sim 0.35 M_\odot$). This is understood because the greater flux of 
neutrinos carries out more energy. However for a finite temperature 
calculation we expect a smaller effect.

A similar analysis was done in \cite{prak95}, where the authors used a 
derivative coupling model with hyperons for the hadron phase and the Bag model 
for the quark matter. They did not obtain any mixed phase for bag values larger
than $B^{1/4}=190$ MeV in contrast with the present work. Moreover, the
maximum masses shown in \cite{prak95} ($\sim 1.6 M_\odot$) and the differences
between maximum masses in neutrino rich and neutrino poor stelar matter are
lower than in our calculations. 

\begin{figure}
\begin{tabular}{cc}
\includegraphics[width=5.cm]{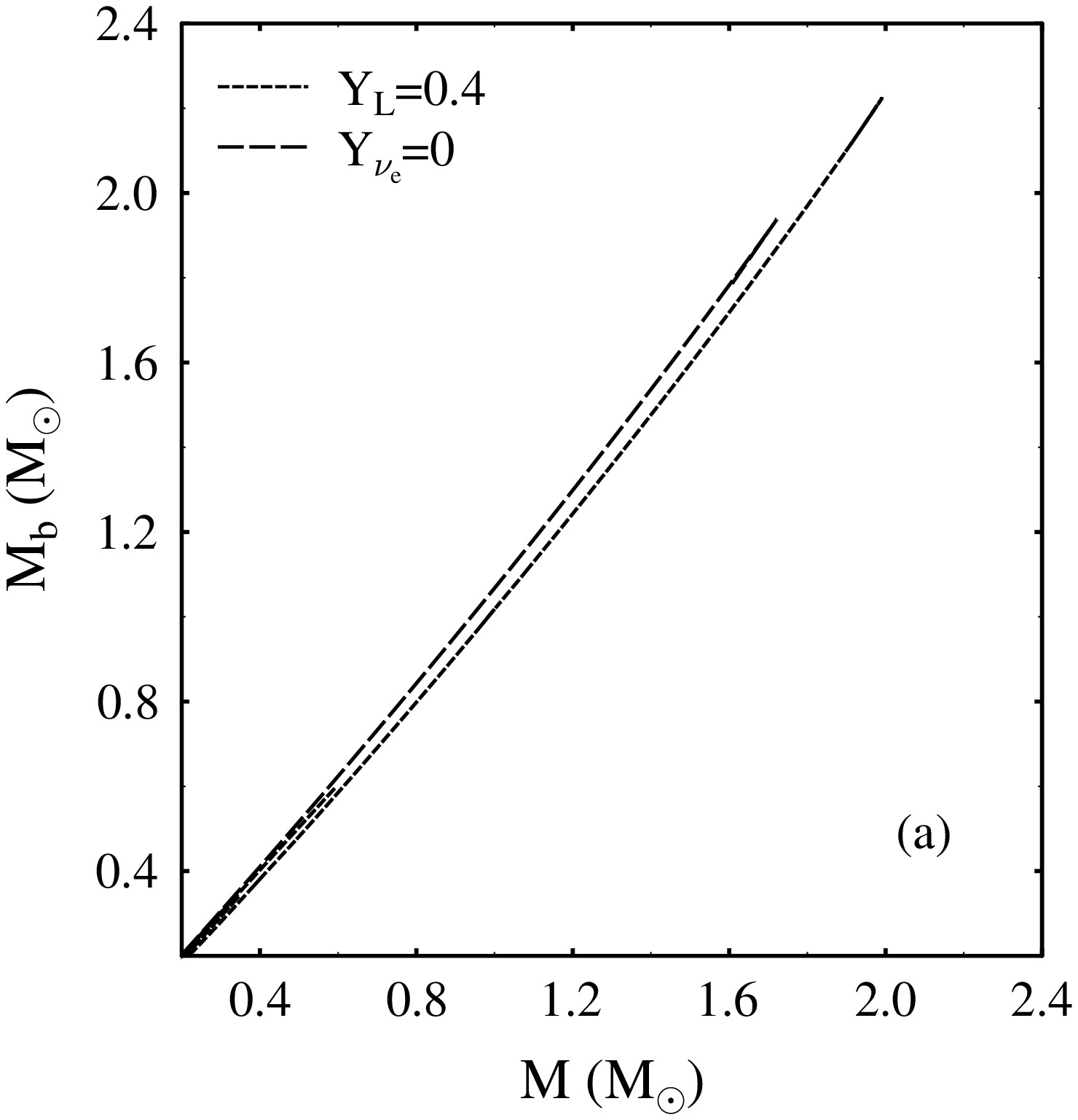} 
&\includegraphics[width=5.cm]{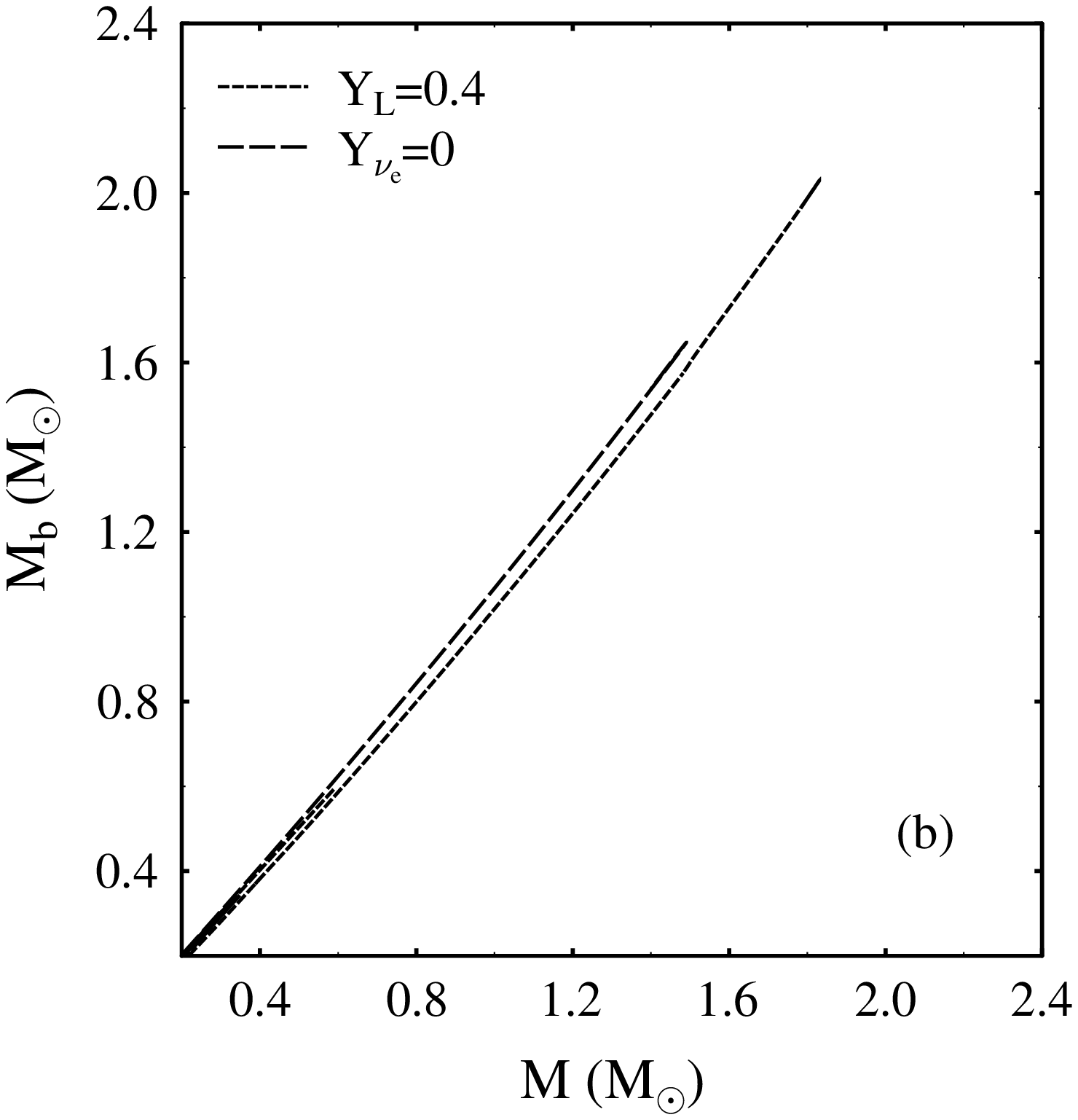}
\end{tabular}
\caption{Baryonic mass versus gravitational mass
 obtained from the EoS with QMC plus (a) UQM (b) CFL.}
\label{massa}
\end{figure}

In fig. \ref{massa} we display the baryonic masses versus the 
gravitational masses for both models. It is seen that neutrino trapped EoSs 
give rise to greater gravitational masses for the same baryonic mass. The 
mass difference reflects the binding energy
released during the deleptonization and cooling stage \cite{prak97}. 
The maximum baryonic mass of  the neutrino rich EoS are larger than the 
neutrino poor. This will lead to a blackhole formation during the 
leptonization period for stars with baryonic masses greater than  the maximum 
baryonic mass of  the neutrino poor EoS. 
This behavior has been encountered in other EoS which include strange matter, 
namely hyperon and/or quark matter \cite{mp,vbpr}.

In summary, we have investigated the effects of neutrino trapping in the 
properties of neutron stars within the QMC framework, including the 
possibility of hyperon formation and a transition
to an unpaired quark-phase or a CFL phase.
We have concluded that within the QMC model with hyperons for the 
hadronic matter, either the hyperons only occur at very high densities or
in very small amounts at lower densities (e.g. UQM).
Another important point is that the maximum mass of a neutrino
rich neutron star decreases after neutrino diffusion leading to the formation 
of a low mass black-hole. This mass reduction is smaller for a quark phase 
described within an unpaired quark phase than a CFL phase. If the quark phase 
is in a CFL state a large fraction of neutrinos is
expected in the mixed and quark phases which will carry away more energy as 
they diffuse out. We also point out that
the amount of neutrinos present in the CFL phase is almost
the double in comparison with the amount found in the UQM phase.
At finite temperature the effect will not be so strong and a 
self-consistent finite temperature calculation should be performed. We have 
also seen that the mass reduction of the maximum mass stars  during  to
neutrino diffusion  is smaller within a QMC formalism than in a NLWM formalism.

This work was partially supported by CNPq (Brazil), CAPES(Brazil), 
GRICES(Portugal) under project 100/03 and FEDER/FCT (Portugal) under the 
project POCTI/35308/FIS/2000.

\end{document}